\documentstyle[twoside,epsfig,fleqn,espcrc2]{article}
\def\ra{\rightarrow}

\newcommand{\AmS}{{\protect\the\textfont2
  A\kern-.1667em\lower.5ex\hbox{M}\kern-.125emS}}

\title{Atmospheric Neutrino Physics with the MACRO detector}

\author{M.Spurio\address{Dipartimento di Fisica dell'Universit\`a and INFN,
        40126 Bologna, Italy\\
        e-mail: spurio@bo.infn.it}
        for the MACRO collaboration}
       
\begin{document}

\begin{abstract}
We present the  measurement of the the flux and angular distribution
of atmospheric $\nu_\mu$
using the MACRO detector. Three different event topologies are
detected in two different energy ranges. 
High energy neutrinos ($\overline E_\nu \sim 80\ GeV$) 
via the identification of upward throughgoing muons.
Lower energy neutrinos ($\overline E_\nu \sim 4\ GeV$) 
via the upgoing stopping and partially contained downgoing muons
(ID+UGS), or via the partially contained  upgoing muons (IU). 
The measured flux is reduced with
respect to the predictions. 
For the high energy sample, globally the flux 
reduction is $0.74\pm 0.054_{stat+sys} \pm 0.12_{th}$; the reduction
varies with the zenith angle.
The ratio of measured to expected events is almost constant 
with the zenith angle for the low energy events, and is 
$0.57 \pm 0.08_{stat+sys} \pm 0.14_{theor}$ for the IU sample, and
$0.71 \pm 0.09_{stat+sys} \pm 0.17_{theor}$ for the (ID+UGS).
All the data sets are consistent within a scenario of neutrino
oscillations, with maximum mixing and
$\Delta m^2 \sim 10^{-3}\div 10^{-2} \ eV^2$. 
\end{abstract}

\maketitle

\section{INTRODUCTION}
The interest on the atmospheric neutrinos has 
grown up in the last year, after the Neutrino '98 Conference in Takayama, 
Japan. New, higher statistic 
data have been presented there by the Soudan 2 \cite{sou98},
MACRO \cite{ronga98}, and SuperKamiokande (SK) \cite{sk98} collaborations. 
The measured flux of muons induced by atmospheric 
$\nu_\mu$ shows
a reduction with respect to the expectation, which depends on the
neutrino energy and direction. For $\nu_e$ induced  electrons 
there is no strong deviation from the prediction. 
The three experiments
explain the $\nu_\mu$ reduction 
in terms of neutrino oscillations, 
with maximum mixing and $\Delta m^2$ few times
$10^{-3}\ eV^2$, confirming the early results. In fact,
in the simplest scenario of two flavor oscillations,
the survival probability of a pure $\nu_\mu$ beam is:
\begin{equation}
P(\nu_\mu \rightarrow \nu_\mu) = 
1- sin^2 2\theta\ sin^2 ( { {1.27 \Delta m^2 \cdot L}\over {E_\nu}})
\end{equation}
$\Delta m^2$ is the mass difference of the two neutrino mass states, $\theta$ is
the mixing angle, $E_\nu$
the neutrino energy and $L$ the path length from the 
production point to the detector. $L$ can be estimated through  
the neutrino arrival direction $\Theta$. For 
upgoing neutrinos, as the zenith angle $\Theta$ changes, 
$L \sim 2R_\oplus\cdot cos\Theta$ ($R_\oplus$ is the Earth radius), while
$\L$ is only few tens of kilometers for downgoing neutrinos.

Atmospheric neutrinos are detected in the
SK water Cherenkov detector via their interaction
with water  nuclei. Three different classes of events are 
defined (with increasing average energy of the parent neutrino): 
fully contained events (FC), partially contained events (PC)
and upward-going muons. Electron neutrinos are also identified in the FC
sample. The ratio of muons 
to electrons normalized to the respective
Monte Carlo predictions enhances the anomaly.

The Soudan 2 results support the oscillation hypothesis
by measuring atmospheric $\nu_\mu$ and $\nu_e$ interactions in the
(roughly) same energy region of SK. A different detection technique
(drift chamber calorimeter) is used in this case.

Here we present the MACRO results on the measurement of
the atmospheric neutrino flux in the energy region from a few
GeV up to a few TeV. 
In this case,  a completely different experimental technique
is used. The flux of $\nu_\mu$ is inferred from the 
measurements of upward throughgoing muons produced
via charged current $\nu_\mu$ interactions
in a large rock volume below the detector. 
The muon can travel up to the apparatus, and if the
residual energy is at least 1 GeV, it is detected as an upgoing muon.
The average energy of
the parent neutrino for these events is $\sim 80\ GeV$, 
one or two order of magnitude larger than that of FC and PC Superkamiokande
events. 
The muon can be detected in MACRO as an upgoing stopping particle,
if its residual energy is between $0.3 \div 1\ GeV$.
In this case,
neutrino interaction happens few meters below the apparatus, and the
average parent neutrino energy is of the order of few GeV.
In addition to the stopping muons, MACRO measures the flux of lower energy
($E_\nu \sim 4\ GeV$) neutrinos through the detection of (mainly)
$\nu_\mu$ interactions inside the apparatus.

\section{MACRO AS $\nu_\mu$ DETECTOR}

\begin{figure}[h]
\centerline{
\epsfig{figure=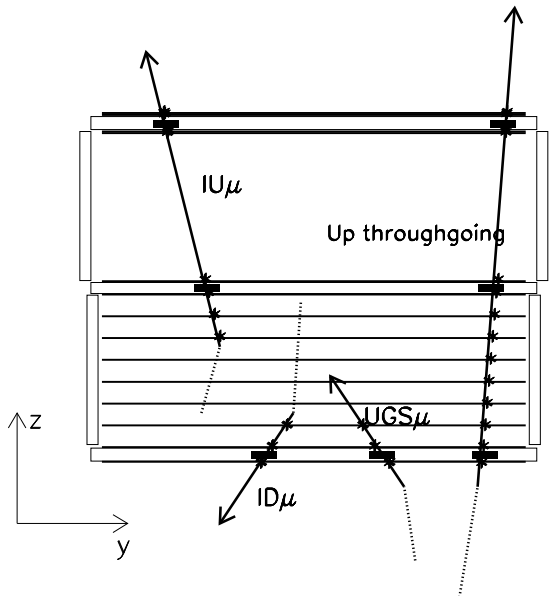,height=7cm} }
\vspace{-0.8cm}
\caption {Sketch of different event topologies 
induced by muon neutrino interactions in or around MACRO. 
The dark boxes represent the
liquid scintillator hits (measurement of time and $dE/dx$), 
while the stars represent 
the hits on the tracking system.}
\label{fig:topo}
\end{figure}

\begin{figure}[h]
\centerline{
\epsfig{figure=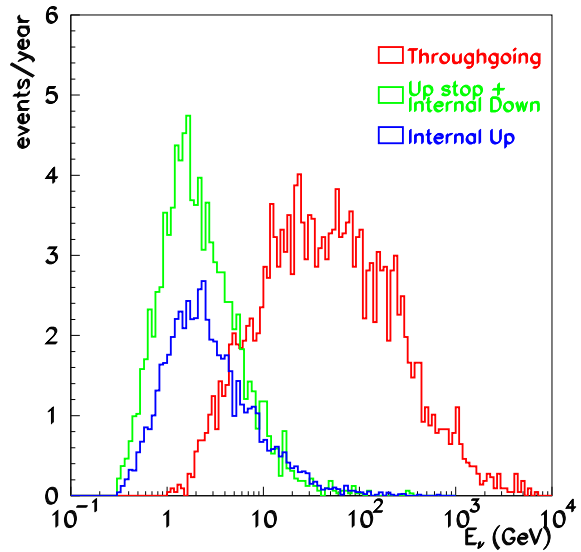,height=80mm} }
\vspace{-0.8cm}
\caption {Expected distribution (for one year of data) of the parent 
$\nu$ energy for the three event topologies in MACRO.}
\label{fig:entop}
\end{figure}

The MACRO detector
is a large rectangular box (76.6~m~$\times$~12~m~$\times$~9.3~m)
whose active detection elements are planes of limited 
streamer tubes for tracking
and liquid scintillation counters for fast timing.
It is located at the Gran Sasso Laboratory, with a minimum rock
overburden of $3150\ hg/cm^2$.
The lower half of the detector is filled
with trays of crushed rock absorber alternating with streamer tube
planes, while the upper part is open. 
The angular resolution for muons achieved by the streamer tube system is
better than $1^\circ$. The
time resolution of each liquid scintillation counter is about $0.5\ ns$.
Fig. \ref{fig:topo}  displays
the different kinds of measured neutrino events.
The {\bf up throughgoing} muons come from $\nu_\mu$
interactions in the rock below the detector. The muon crosses the
whole detector and the flight direction is determined by
time-of-flight ({\it t.o.f.}) measurement.
The flux of lower energy $\nu_\mu$ is studied by the
detection of $\nu_\mu$ interactions inside the apparatus;
the partially contained upgoing events ({\bf IU}) are tagged
with {\it t.o.f.}. The partially contained downgoing events ({\bf ID}) 
and upward going stopping muons  ({\bf UGS}) are 
identified via topological constraints.
Fig. \ref{fig:entop}  shows 
the distribution of the parent neutrino energy for the three event
topologies detected by MACRO.
The data presented here come mainly from the running period with the
full MACRO detector (which started acquisition in April 1994), 
up to February 1999, corresponding to an effective 
live-time of $4.1\ years$.

\vskip -1cm

\section{UP THROUGHGOING MUONS}
\begin{figure} [htb]
\centerline{
\epsfig{file=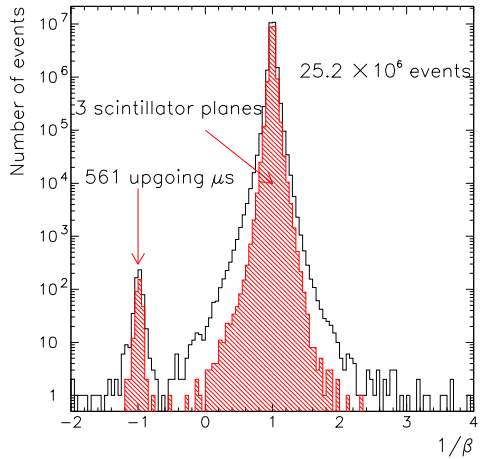,height=7cm} }
\vspace{-1.3cm}
\caption { $1/\beta$ distribution (data with the full detector); the
upgoing muons are centered at $1/\beta =-1$. 
The shaded distribution is for the subset of events with
three scintillator layers.}
\label{fig:sbeta}  
\end{figure}

At the depth of the Gran Sasso Laboratory, 
we expect $\sim 5\times 10^4$ downgoing
atmospheric muons for each neutrino-induced $\mu$ (for which we expect an
up/down symmetry).
For this reason, the identification of neutrino-induced $\mu$'s relies
on the  measurement of the direction that muons travel through MACRO.
For each detected muon the experimental parameter
$1/\beta = {{c\cdot (T_1 - T_2)}\over {D}}$  is
evaluated.  $T_1\ (T_2)$ is the time measured in the lower (higher) 
scintillation
counter and $D$ the path length between the counters. 
The downgoing muons are expected in the 1/$\beta$ region near +1, 
while upgoing muons at 1/$\beta$ near -1.
To remove accidental background events, the 
position along the scintillator counter must agree
within $\pm$70 cm with the position indicated 
by the streamer tube track.
Fig. \ref{fig:sbeta}  shows the distribution of 
$1/\beta$ for muons collected from April 1994.
There are 561 upgoing muons
in the range $-1.25 < 1/\beta < -0.75$. We combine
these data with additional 81 events collected before 1994,
for a total of 642 upgoing events.
In the total data set there are: $12.5 \pm 6$ estimated
background events due to misidentification of downgoing muons;
$10.5 \pm 4$ background events from upgoing
charged particles produced by downgoing muons in the rock near MACRO 
\cite{macrouppi}; $12 \pm 4$ internal events from
interactions of neutrinos in the very bottom layer of MACRO
scintillator. Removing the background and the internal events, the
number of upward throughgoing muons is 607.

\subsection{Monte Carlo expectation}

For the simulation, we have used the Bartol neutrino flux
\cite{agrawal} and the DIS parton
distribution set \cite{gluck} for the neutrino cross-sections.
The propagation of muons
to the detector has been done using  the energy
loss calculation \cite{lohmann} for standard rock.
The total uncertainty on the expected flux of muons
adding in quadrature the errors from neutrino flux, cross-section
and muon propagation is $\pm17\%$.
This theoretical error in the prediction
is mainly a scale factor that doesn't change the shape of the angular
distribution.
The number of expected events is
824.6, giving a ratio of the observed number of events to the expectation
of $0.74\pm0.031_{stat} \pm 0.044_{sys} \pm 0.12_{theo}$.
Fig. \ref{fig:flux}  shows the zenith angle distribution of the measured
flux of up throughgoing muons (all MACRO data),
compared to Monte Carlo expectation.

\subsection{Interpretation of the result}

\begin{figure} [tbh]
\centerline{
\epsfig{file=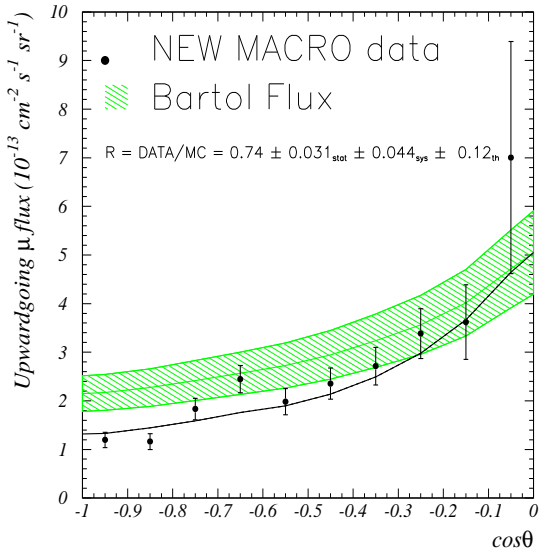,height=7cm} }
\vspace{-0.8cm}
\caption{Measured flux of upward throughgoing muon $vs.$ the cosine of
zenith angle $\Theta$. The 17\% 
uncertainty (shadow) is almost a constant corrective
factor to the central value, being the error on the shape almost negligible.
The lower line shows the prediction assuming two-flavors neutrino 
oscillations (see text).}
\label{fig:flux}
\end{figure}
                                               
\begin{figure}[htb]
\centerline{
\epsfig{figure=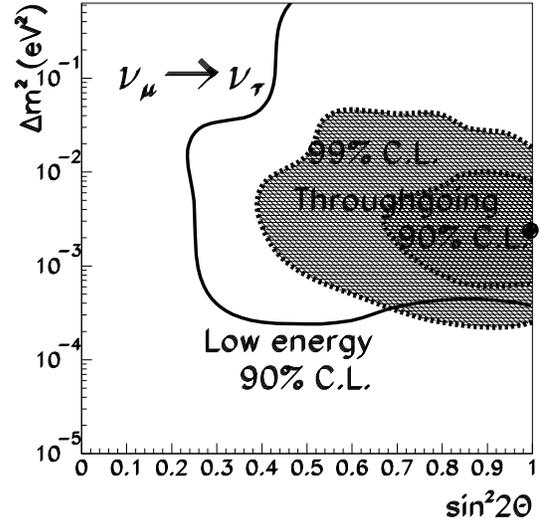,height=7cm} }
\vspace{-0.8cm}
\caption{Allowed  contours (and best fit point) at 90\% and 99\% C.L. 
        (darker region)
        assuming $\nu_\mu \ra \nu_\tau$ oscillations. 
        The 90\% C.L. contour obtained from the 
        lower energy neutrino events (IU and ID+UGS) is also shown.}
\label{fig:allow}
\end{figure}

We interpreted the reduction on the detected number of events
and the deformation of the zenith angle distribution 
as a consequence of $\nu_\mu$ disappearance.
In the scenario described by eq. 1,
relatively fewer events are expected near the vertical 
$(cos\Theta=-1)$ than near the horizontal $(cos\Theta=0)$,
due to the longer path length of neutrinos from production to observation.
The shape of the angular distribution 
has been tested with the hypothesis of no-oscillation.
We found a $\chi^2/d.o.f =22.9/8$, or a probability of 0.35\%.
To test the  $\nu_{\mu} \rightarrow\nu_{\tau}$
oscillation hypothesis, we evaluate  the
independent probability for
obtaining the number of observed events and the angular distribution
for various oscillation parameters. 
The maximum of the $\chi^2$ probability in the physical region of 
the oscillation parameters is  36.6\%, corresponding to
$\Delta m^2 =2.5\times 10^{-3} eV^2$ and maximum mixing.
Fig. \ref{fig:allow}  shows the
confidence regions at the 90\% and 99\% C.L. 
in the parameter space $(\sin^2 2 \theta , \Delta m^2)$ for
$\nu_{\mu}  \rightarrow\nu_{\tau}$ oscillations, based on
application of the Monte Carlo prescription of \cite{feldman}.
The effect of oscillations (using the best fit point values)
is also shown in Fig. \ref{fig:flux}. 
The angular distribution of events {\it vs.} $cos\Theta$ strongly 
disfavours the 
hypothesis of no-oscillations; however, there is a structure (near
$cos\Theta \sim -0.65$) which is unexpected also in the case of oscillations.
The excess of events in this bin is still consistent with 
a statistical fluctuation, but it could be a hint for a more complex scenario.

\subsection{Checks on systematics}
Possible systematic effects have been studied and have been shown to be too 
small to explain the observed anomalous shape in the zenith angle distribution.
The detector acceptance is well understood (from a large sample of downgoing
muons), in particular near the vertical direction
where the largest deviation compared
to the Monte Carlo expectation without oscillations is observed.

\begin{figure}[tbh]
\centerline{
\epsfig{figure=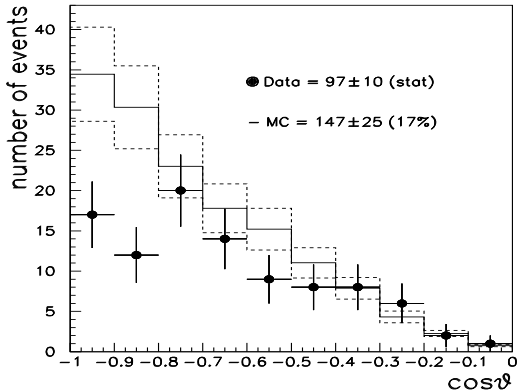,height=6cm,width=7cm} }
\vspace{-0.8cm}
\caption{Zenith angle distribution of the upgoing muons data (points)
compared with the Monte Carlo predictions without oscillations
for the events measured with the system dedicated for the 
search of stellar collapses ($\sim 1.7\ y$ live time)}
\vskip -1cm
\label{fig:phrase}
\end{figure}

An important check has been performed using the system designed for the
detection of bursts of neutrinos from stellar collapse. This system has
a completely separated acquisition system and redout electronics. An analysis
has been done using the TDC's of such system and the events having three
scintillation counters. With three counters is possible to achieve a good
rejection of downgoing muons using only the information from this
TDC system and without tracking. The angular distribution of the 
up throughgoing event sample identified in this way
(presented in Fig. \ref{fig:phrase}) confirms the main
feature of Fig. \ref{fig:flux}. 
In the figure some events collected when the stellar collapse system 
was on and the main system off are included.
During the period of common activity,
only one event has escaped from the detection of the 
standard analysis, while  
we expect to miss 0.6 events due to dead time of the 
the streamer tube/liquid scintillation systems used by the standard analysis.

\section{EVENTS FROM LOWER ENERGY $\nu_\mu$}

\subsection{Partially contained upgoing $\mu$ (IU)}

The detection methods for the IU are
similar to that of throughgoing $\mu$'s
(time-of-flight measurement,
plus a track reconstructed
in the streamer tubes system), and on topological
criteria for the identification of a interaction vertex inside the apparatus.
To reject fake semi-contained events entering from a
detector crack, the extrapolation of the track in the lower part
of the detector must cross and not fire at least three streamer tube planes
and one scintillation counter. The above conditions, tuned on the
Monte Carlo simulated events,
account for detector inefficiencies and reduce
the contribution from upward throughgoing muons which appear
like semi-contained to less than $\sim 1 \%$.
We evaluated that $5$ events are due to an uncorrelated background.
After the background subtraction, $116$ events are classified as $IU$ events.
              
\subsection{Partially contained downgoing and up stopping events (ID)}
The ID+UGS events cross only one liquid scintillator layer
and are identified by means of topological criteria.  The lack
of timing information prevents to distinguish between the two sub samples.
From MC (sec. 4.3.), an almost equal number of $UGS$ and $ID$ events 
are expected.
A software selection rejects throughgoing events, and searches
for a contained track crossing the bottom layer of the scintillation counters.
The selection conditions for the event vertex (or $\mu$ stop point)
in the detector are symmetrical to those for the $IU$ search.
879 events are accepted by the selection.
Some of them are wrongly tracked or bending atmospheric muons which 
entered from a detector crack.
To reject fake events, a visual scan was performed.
Two physicists scanned twice the
real data randomly merged with the simulated events.
At the end of the scan procedure, 200 real events are accepted as 
ID or UGS, with 95\% of the Monte Carlo simulated events.
The main background source are upward going charged pions
induced by interactions of atmospheric muons in the rock
around the detector. 
$7.2\pm 2.3 $ background events have been evaluated 
using a full simulation which is based on our measurement \cite{macrouppi}.

\subsection{Monte Carlo}
Also the expected low energy event rates have  been evaluated 
with a full Monte Carlo simulation.
The interaction of atmospheric $\nu_e$ and $\nu_\mu$ \cite{agrawal} were 
simulated in a large volume (with $175\ kton$  total mass), 
including the experimental hall and the detector.  
Because of the lower $\nu$ energies, 
the cross sections from \cite{lipari} are used, which include the contribution
of the exclusive channels of quasi-elastic scattering and single-$\pi$
production. The total theoretical uncertainty on $\nu$ flux  
and  cross section at these energies is $\sim 25\%$.
The detector response has been simulated
using a GEANT based program, and events are processed in the
same analysis chain as the real data. The parameters of
the streamer tube and scintillator systems have been chosen to
reproduce the real average efficiencies.
A 10\% systematic error  is evaluated
from the simulation of detector response,
data taking  conditions, analysis algorithm efficiency,
mass and acceptance of the detector.
As shown in Fig. \ref{fig:entop}, the energy spectra of
parent neutrinos for the 
{\it IU} and {\it ID+UGS} events are similar, with equal average energy.
87\% of IU and (ID+UGS) detected events are induced by $\nu_\mu$-CC 
interactions. The remaining 13\%, from $\nu_e$-CC and NC interactions
(with different percentage for the two data sets).

\subsection{Results from the low energy events}

\begin{figure}[tbh]
\centerline{
\epsfig{figure=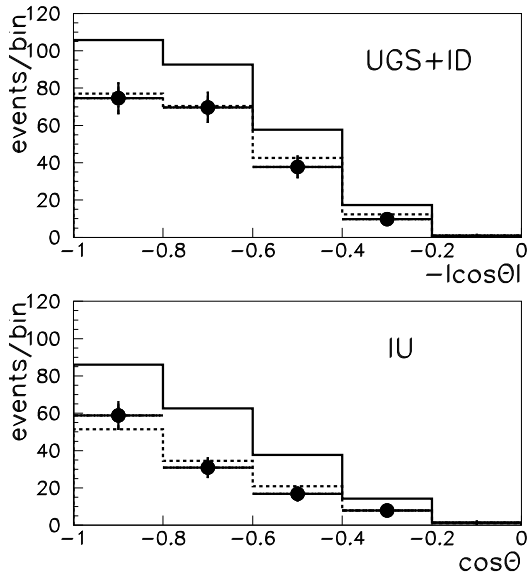,height=76mm} }
\vspace{-0.8cm}
\caption {Cosine of the zenith angle ($\Theta$) distribution for 
(ID+UGS) and IU events.
The background-corrected data points (black points with error bars)
are compared with the Monte Carlo
expectation assuming no oscillation (full line) and two-flavour
oscillation (dashed line) using maximum mixing and
$\Delta m^2=2.5 \times 10^{-3}\ eV^2$.}
\label{fig:lebo}
\end{figure}

Fig. \ref{fig:lebo}  shows the zenith angle distribution of 
the $IU$ and $UGS + ID$ data samples,
with the Monte Carlo predictions.
The data are within errors consistent with a constant
deficit in all bins with respect to the Monte Carlo expectations.
The ratios of the number of observed to expected events are
$R_{ID+UGS} = ({{Data}\over {MC}})_{ID+UGS} =
0.71 \pm 0.05_{stat} \pm 0.07_{syst} \pm 0.17_{theor}$ and
$R_{IU} = 0.57 \pm 0.05_{stat} \pm 0.06_{syst} \pm 0.14_{theor}$.

If the event deficit is due to an overall theoretical overestimate of
the neutrino flux and/or cross sections, it is expected 
${\cal R}=R_{IU}/ R_{ID+UGS}=1$.
The theoretical and systematic 
errors are largely reduced (to 4\% and 5\%, respectively) 
if the ratio of ratios is considered. 
The partial uncertainty cancelation comes from the fact that the 
energy spectra of the two topologies are alike, and because of
the symmetry in the detector acceptance.
We measured ${\cal R}=0.80\pm 0.09_{stat}$; the statistical error is
the dominant one in this quantity.
The probability to obtain a ratio so different from the expected one is 5\%
(taking into account the non-gaussian shape of the uncertainty),
no matter which neutrino flux and neutrino cross sections are used
for the predictions.

The alternative hypothesis for such a reduction is
$\nu_\mu \rightarrow \nu_\tau$ oscillations with maximum mixing and
$\Delta m^2 \sim 10^{-3} \div  10^{-2}\ eV^2$.
In this $\Delta m^2$ interval the number
of muons induced by the interaction of upgoing neutrinos
({\it i.e.} IU and UGS events, for which
$L\simeq 13000\ km\ , \overline E_\nu \simeq 4\ GeV$)
is reduced by a factor of two, while
almost no reduction is expected for ID events ($L$= few $km$).
The expected (ID+UGS) event rate is 3/4 of the no-oscillations
expectation (neglecting the contribution of
$\nu_e$  and NC interactions).
For larger $\Delta m^2$, also the ID events are reduced, so both the
(ID+UGS) and IU event rates are 1/2 of the no-oscillations
expectation. For smaller $\Delta m^2$, the bins of the
zenith distributions (Fig. \ref{fig:lebo})
are differently affected by the oscillations. 
As the low-energy events are not particularly sensitive to the
$\Delta m^2$ interval $\sim 10^{-3} \div  10^{-2}\ eV^2$
as a test point in Fig. \ref{fig:lebo} we used the best fit values from the
high energy sample.

Finally, the $\nu_\mu \rightarrow \nu_\tau$ hypothesis  simply explain
the different reduction on the total number of events
observed in the (ID+UGS) and IU data sets. 
In case of $\nu_\mu$ disappearance, for the  (ID+UGS)
the reduction with respect to the case of no-oscillations is 0.76, to
be compared with the measured value of $R_{ID+UGS} = 0.71$.
For the IU events, it is 0.57, to be compared with 
$R_{IU} = 0.57$

We estimated the most likely values 
on a $(sin^2 2\theta, \Delta m^2)$ grid
using a $\chi^2$ comparison of data and Monte Carlo, based on the 
prescription of \cite{feldman}.
The data were binned in 4 zenith angle bins for the IU events, 
4 zenith angles for the ID+UGS events,
the ratio ${{IU}\over {ID+UGS}}$ and the overall normalization.
The maximum of the $\chi^2$ probability
(97\%) occurs at $sin^22\theta =1.0$
(inside the physical region);
this value of the $\chi^2$ probability is almost constant in the interval
$\Delta m^2 = 10^{-3}\div 2.\times 10^{-2}\ eV^2$.
Fig. \ref{fig:allow}
shows the  contour 90\% confidence level for the low energy events;
the allowed region is consistent with that
obtained using the higher energy sample of neutrino-induced upward
throughgoing muons.

\section{CONCLUSIONS}
MACRO measures three different data sample of events induced by 
atmospheric neutrinos. All the data sets show a deficit of the
measured number of events with respect to the predictions based on the 
Bartol flux and the absence of neutrino oscillations.

For the upward throughgoing sample, apart from the reduction in number,
the shape of the angular distribution is modified as expected
in case of $\nu_\mu$ disappearance. There is however an unexpected structure 
near $cos\Theta \sim -0.65$), which is still consistent as a 
a statistical fluctuation in the scenario of neutrino oscillations.

For the low energy neutrinos, two samples are measured. The IU events
are induced by upgoing neutrinos, while the ID+UGS events are 50\% from
neutrino from above (ID). 
The measured number of events is  below the
expectations, but the reduction is different for the two data set.
Because the parent neutrinos for the IU and the ID+UGS events have a quite 
similar energy spectrum, an overall reduction of the number 
of neutrino-induced muons has a low probability (5\%) to explain
the two observed different deficit.
This effect is explained with larger probability
($\sim$ 97\%) by the hypothesis of muon neutrino oscillations
with maximum mixing and  $\Delta m^2 = 10^{-3}\ \div 2\times 10^{-2}\ eV^2$.

In Table 1 is presented a summary of the number of detected
events, the number of the expected ones in case of no-oscillation
and for $\nu_\mu \ra \nu_\tau$ oscillations with 
maximum mixing and $\Delta m^2=2.5 \times 10^{-3}\ eV^2$.
The up throughgoing sample has the higher sensitivity
to the oscillation parameters.
For this sample, the detection technique and the average energy
of the parent neutrino are completely different from the SK and Soudan 2 ones.
However, the allowed region 
in the oscillation parameter space largely overlap
that of SK and Soudan 2, and also the best fit parameters are similar to the
SK one.

\vskip -0.8cm
\begin{table}[htb]
\begin{center}
\begin{tabular}
{cccc}\hline
 & {\bf Data} & \multicolumn{2}{c}{\bf MC (Bartol flux)}  \\ \cline{3-4}
 &            & No osci & With osci \\ 
 & - Bckg & $\pm_{sta+sys}$ & $0.0025 eV^2$       \\ \hline
Up Through  & 607 & $824\pm 56$ & $585$ \\ 
IU                 & 116 & $202\pm 22$ & $115$ \\
ID+UGS               & 193            & $273\pm 30$    & $209$ \\ \hline
\end{tabular}
\end {center}
\caption{Summary for the MACRO measurements of the
atmospheric $\nu$ flux. For the three event topologies, the
number of detected events (first column) is compared with the expectation
in case of no-oscillations, and for $\nu_\mu \ra \nu_\tau$ oscillations with 
maximum mixing and $\Delta m^2=2.5 \times 10^{-3}\ eV^2$
(best fit point for the high energy set).}
\end{table}

\end{document}